\documentclass[11pt,english,epsf]{scrartcl}
\usepackage{lmodern}
\usepackage[T1]{fontenc}
\usepackage[latin9]{inputenc}
\usepackage{geometry}
\usepackage{amsmath}
\usepackage{amssymb}

\newtheorem{theorem}{Theorem}

\newcommand {\dfn} {\stackrel{\Delta} {=}}

\newcommand{\lea}{\stackrel{\mbox{(a)}}{\le}}

\newcommand {\reals} {{\rm I\!R}}

\newcommand {\bq} {\mbox{\boldmath $q$}}
\newcommand {\bpi} {\mbox{\boldmath $\pi$}}

\newcommand {\bs} {\mbox{\boldmath $s$}}
\newcommand {\ts} {\tilde{s}}
\newcommand {\tx} {\tilde{x}}

\newcommand {\bw} {\mbox{\boldmath $w$}}
\newcommand {\bx} {\mbox{\boldmath $x$}}
\newcommand {\by} {\mbox{\boldmath $y$}}

\newcommand {\bV} {\mbox{\boldmath $V$}}

\newcommand{\calC}{{\cal C}}
\newcommand{\calD}{{\cal D}}
\newcommand{\calE}{{\cal E}}

\newcommand{\calQ}{{\cal Q}}

\newcommand{\calS}{{\cal S}}
\newcommand{\calT}{{\cal T}}
\newcommand{\calU}{{\cal U}}

\newcommand{\calX}{{\cal X}}
\newcommand{\calY}{{\cal Y}}

\allowdisplaybreaks

\begin{document}
\thispagestyle{empty}

\title{On Zero--Rate Error Exponents of Finite--State Channels
with Input--Dependent States}
\author{Neri Merhav}

\date{}
\maketitle

\begin{center}
Department of Electrical Engineering \\
Technion - Israel Institute of Technology \\
Technion City, Haifa 32000, ISRAEL \\
E--mail: {\tt merhav@ee.technion.ac.il}\\
\end{center}
\vspace{1.5\baselineskip}
\setlength{\baselineskip}{1.5\baselineskip}

\begin{center}
{\bf Abstract}
\end{center}
\setlength{\baselineskip}{0.5\baselineskip}

We derive a single--letter formula for the zero--rate reliability (error
exponent) of
a finite--state channel whose state variable depends
deterministically (and recursively) on past channel inputs, where the code
complies with a given channel input constraint.
Special attention is
then devoted to the important special case of the Gaussian channel with inter-symbol
interference (ISI), where more explicit results are obtained.

\vspace{0.2cm}


\setlength{\baselineskip}{2\baselineskip}
\newpage

\section{Introduction}

The concept of the reliability function of a channel is almost as old as
information theory itself. The first to show that below capacity, the
probability of error decays exponentially with the block length, for a
sequence of good codes, was Feinstein \cite{Feinstein55} in 1955. 
Already in the same year, Elias \cite{Elias55} derived the random coding
bound and the sphere--packing bound, and he observed that they exponentially
coincide at high rates, for the cases of the binary symmetric channel
(BSC) and the binary erasure channel. Six years later, Fano
\cite{Fano61}, derived the random coding exponent, $E_{\mbox{\tiny r}}(R)$,
and heuristically also the sphere--packing bound for the general discrete memoryless
channel (DMC). In 1965, Gallager \cite{Gallager65} improved on $E_{\mbox{\tiny
r}}(R)$ at low rates by the idea of expurgation of randomly selected codes. 
In 1967, Shannon, Gallager, and Berlekamp, published their celebrated two--part paper
\cite{SGB67a}, \cite{SGB67b}, where they derived the classical lower bounds on the error
probability for general
DMC's: the sphere--packing bound, the zero--rate bound, and the straight--line
bound, that improves on the sphere--packing bound at low rates, using the
zero--rate bound. 

In the realm of channels with memory, the most popular model dealt with, in
this context, has
been the model of a finite--state channel (FSC) and some of its special cases. 
The channel coding theorem for FSC's was proved 
by Blackwell, Breiman and Thomasian \cite{BBT58} in 1958. 
The random coding exponent for FSC's was derived by Blackwell
\cite{Blackwell61} in 1961, Yudkin \cite{Yudkin67} in 1967,
and further developed by Gallager in his book \cite[Section 5.9]{Gallager68},
especially for the case where the state is known at the receiver.

Ever since these early days of information theory, there has been a
vast amount of continued work around error exponents and reliability functions, most
notably, for memoryless channels (both discrete and continuous), but also 
(albeit, much less) for various models of 
channels with memory (FSC's included), both in the presence and in the absence of
feedback. For the latter category, see, e.g.,
\cite{Arikan02}, \cite{CYS07}, \cite{CYS09}, \cite{EZ01}, \cite{Lapidoth93},
\cite{NGW06}, \cite{PV95}, \cite{SKN82},
\cite{WA05}, \cite{Ziv85}, \cite{ZAC05} and references therein, for a non--exhaustive list of
relevant works from the
last three decades.

In this paper, our focus is on the zero--rate reliability of channels
from  a subclass of the {\it FSC's with input--dependent states}
(without feedback), namely, finite--state channels where the
state variable, which designates the memory of the channel,
evolves deterministically in response to past channel
inputs, as opposed to the more general channel model, where the state evolves stochastically
in response to both past inputs and outputs. For a 
finite input alphabet, this subclass of FSC's is still
general enough to include the important model of the
inter-symbol interference (ISI) channel, among some other models.

Our primary motivation for studying the zero--rate reliability for these
channels is in order to identify and characterize, by means of single--letter
formulas, the relevant distance
metrics and the maximum achievable minimum distance between codeword pairs 
under this metric, in analogy to the Hamming distance for the BSC, the
Euclidean distance for the Gaussian memoryless channel, and the Bhattacharyya
distance for a general DMC. A secondary motivation is that once the zero--rate reliability is known and
the sphere--packing bound is known, at least for some positive rate, one can
obtain a simple bound at all rates using the straight line in between, by
using a straightforward extension of
\cite[Theorem 1]{SGB67a} (see also \cite[Theorem 3.8.1]{VO79}) to channels
with memory. For example,
even if the sphere--packing exponent is not available, but
the capacity $C$ of the channel is known (or at least we have an upper bound
for it), then we know that the sphere--packing bound at rate $C$
vanishes, and we can safely use this theorem to connect the above--mentioned straight line to
the point $(C,0)$ in the plane of reliability vs.\ 
rate.\footnote{This is supported by the fact for any code of rate
just above $C+\lambda$ ($\lambda > 0$), the probability of list--error, for
an exponential list size of rate $\lambda$, must be bounded away from zero,
as can easily be seen from a simple extension of Fano's inequality to list
decoding.} This bound can be reasonably good at least for low rates. 

Our main result, in this paper, is an exact single--letter characterization of the zero--rate
reliability (or the maximum achievable minimum `distance') for FSC's with
input--dependent states, and codes that must conform with a given input
constraint. More explicit results are provided in the Gaussian case
with inter-symbol interference (ISI), which
will be treated in some detail later in the paper. 

\section{Preliminaries}

Before addressing FSC's, we begin with some preliminaries on the zero--rate
reliability of a DMC. Let us define
\begin{eqnarray}
\label{e0def}
\calE_0^+&\dfn&\lim_{R\downarrow
0}\limsup_{n\to\infty}\left[-\frac{\ln P_e(R,n)}{n}\right]\\
\calE_0^-&\dfn&\lim_{R\downarrow
0}\liminf_{n\to\infty}\left[-\frac{\ln P_e(R,n)}{n}\right],
\end{eqnarray}
where $P_e(R,n)$ is the minimum probability of error that can be attained,
for the given channel, by any block code of length $n$ and rate
$R$. Consider a DMC, designated by a matrix of input--output transition
probabilities $\{p(y|x),~x\in\calX,~y\in\calY\}$. Here the channel input
symbol $x$ takes on values in a finite input alphabet $\calX$, whereas the channel
output symbol $y$ takes on values in the output alphabet $\calY$, which may
either be
discrete or continuous.\footnote{We proceed hereafter under the
assumption that $\calY$ is a discrete alphabet, but with the understanding that
in the continuous alphabet case, all probability distributions over $\calY$
are replaced by densities, and accordingly, all summations over $\calY$ should be replaced by
integrals.}
When the channel is fed by a vector
$\bx=(x_1,\ldots,x_n)\in\calX^n$, it outputs a vector
$\by=(y_1,\ldots,y_n)\in\calY^n$ according to
\begin{equation}
P(\by|\bx)=\prod_{t=1}^n p(y_t|x_t).
\end{equation}

For DMC's whose zero--error capacity vanish,
the zero-rate reliability is well--known \cite{SGB67a},
\cite{SGB67b} to be given by\footnote{The zero--rate reliability is more
commonly denoted by
$E_{\mbox{\tiny ex}}(0)$, as it is identified with the expurgated error
exponent at rate zero.
However, since we consider here zero--rate codes only, we will use the
more convenient notation $E_0$, with no
risk of confusion with customary notation concerning the Gallager function and
random coding exponents, as these quantities will not be addressed in this
paper.}
\begin{equation}
\label{E0}
\calE_0^+=\calE_0^-=E_0\dfn\max_{\bq}\left[\sum_{x,x'\in\calX}q(x)q(x')d_{\mbox{\tiny
B}}(x,x')\right],
\end{equation}
where $d_{\mbox{\tiny B}}(x,x')$ is the Bhattacharyya distance function,
defined as
\begin{equation}
d_{\mbox{\tiny
B}}(x,x')=-\ln\left[\sum_{y\in\calY}\sqrt{p(y|x)p(y|x')}\right],~~~~x,x'\in\calX,
\end{equation}
and the maximum is over all possible probability assignments,
$\bq=\{q(x),~x\in\calX\}$, over the input
alphabet. This is the best attainable error exponent for {\it any} code over
$\calX$. 

In the presence of input constraints, the expression (\ref{E0}) may not be
achievable since the optimal codes might violate these constraints. 
For example, suppose that each codeword in the codebook must satisfy
the constraint 
\begin{equation}
\label{inputconstraint}
\sum_{t=1}^n\phi(x_t)\le n\Gamma,
\end{equation}
where $\phi:\calX\to\reals$ is a given function (e.g., $\phi(x)=x^2$) and
$\Gamma$ is a prescribed quantity.
At first glance, it may be tempting to guess that the best achievable exponent
would then be the same as in (\ref{E0}), except that the maximum over $\bq$
should be restricted to
comply with the corresponding single--letter
constraint, that is, $\bq\in\calQ_\Gamma$, where
$\calQ_\Gamma=\{\bq:~\sum_xq(x)\phi(x)\le \Gamma\}$.

It turns out, however, that this is indeed true for some channels, but not in
general. In certain cases, one can do better. The point is that the functional
\begin{equation}
E_0(\bq)\dfn \sum_{x,x'\in\calX}q(x)q(x')d_{\mbox{\tiny
B}}(x,x')
\end{equation}
may not, in general, be concave in $\bq$. This depends on the given (symmetric) matrix of Bhattacharyya
distances, $D=\{d_{\mbox{\tiny B}}(x,x')\}_{x,x'\in\calX}$,
which in turn, depends solely on the channel and the input
alphabet. If $D$ is such that $E_0(\bq)=\bq^TD\bq$ ($\bq$ being thought of as a column
vector), is concave, then $E_0(\bq)$ is the best exponent achievable for codes
with codebooks of composition\footnote{The composition $\bq$ of a fixed
composition code is the empirical distribution of each one of the codewords.}
and hence $\max_{\bq\in\calQ_\Gamma}E_0(\bq)$ is indeed the best achievable exponent under the
aforementioned input constraint. If, however, $E_0(\bq)$ is not concave, one can
improve by taking the upper concave envelope (UCE) of $E_0(\bq)$ (see
\cite[p.\ 191, Problem 21]{CK81}). Accordingly, let us denote
\begin{equation}
\label{uce}
\overline{E}_0(\bq)=\mbox{UCE}\{E_0(\bq)\}\dfn
\max_{\{(\bw,\bV):~\sum_uw(u)v(x|u)=q(x)~\forall x\in\calX\}}
\sum_{u\in\calU}w(u)\sum_{x,x'}v(x|u)v(x'|u)d_{\mbox{\tiny
B}}(x,x'),
\end{equation}
where $\bw=\{w(u),~u\in\calU\}$ is a probability vector of a (time--sharing)
variable $u$, whose alphabet size $|\calU|$ need not exceed $|\calX|$ (as can
easily be shown using the Carath\'eodory theorem \cite[p.\ 310, Lemma 3.4]{CK81}), and
$\bV=\{v(x|u),~u\in\calU,~x\in\calX\}$ is a matrix of transition probabilities
of $x$ given $u$. The input constraint is then accommodated for
$\overline{E}_0(\bq)$, that is, the best attainable exponent is
$\max_{\bq\in\calQ_\Gamma}\overline{E}_0(\bq)$.

When $E_0(\cdot)$ is concave, the operator $\mbox{UCE}\{\cdot\}$ is, of course,
redundant, so it is instructive to know when is this the case.
The concavity of $E_0(\bq)$ over the simplex can easily be checked as follows.
Without loss of generality, let $\calX=\{1,2,\ldots,K\}$, $K=|\calX|$.
On substituting $q(K)=1-\sum_{x<K}q(x)$ into the quadratic form
$E_0(\bq)=\bq^TD\bq$, one ends up with the reduced quadratic form 
$\tilde{\bq}^T\tilde{D}\tilde{\bq}$, where
$\tilde{\bq}=\{q(x),~x=1,2,\ldots,K-1\}$ and $\tilde{D}$ is a
$(K-1)\times(K-1)$ whose $(x,x')$--th entry is $d_{\mbox{\tiny
B}}(x,x')-d_{\mbox{\tiny B}}(x,K)-d_{\mbox{\tiny B}}(K,x')$,
$x,x'\in\{1,2,\ldots,K-1\}$. Thus, $E_0(\bq)$ is concave iff $\tilde{D}$ is
negative semi--define, or equivalently, iff $-\tilde{D}=\{d_{\mbox{\tiny
B}}(x,K)+d_{\mbox{\tiny B}}(K,x')-d_{\mbox{\tiny B}}(x,x')\}$ is positive
semi--definite. We henceforth denote by $\calD(K)$ the class of matrices
$\{D\}$ for which $\tilde{D}$ is negative semi--definite.\footnote{Of course, the choice of the
letter $x=K$ as the one with the special stature here is completely arbitrary.}

It should be pointed out that for some rather important special cases, $D\in\calD(K)$
and hence $E_0(\bq)=\bq^TD\bq$ is concave on the simplex.
For example, if $d_{\mbox{\tiny B}}(x,x')$ is (proportional to) the Hamming distance (which is always
the case, for example, when $K=2$), 
then $-\tilde{D}$ is a matrix whose all diagonal elements are $2$ and all
off--diagonal elements are $1$. The eigenvalues of this matrix are $0$ and $K$ (the former,
with multiplicity of $K-2$) and hence it is positive semi--definite.
As another example, if $d_{\mbox{\tiny B}}(x,x')$ is (proportional to) the
square--error, $(x-x')^2$, which is the case when the channel is Gaussian, 
then the $(x,x')$--th element of $-\tilde{D}$ is
$(x-K)^2+(x'-K)^2-(x-x')^2=2(x-K)(x'-K)$, which is obviously positive semi--define,
with eigenvalues $2\sum_{x<K}(x-K)^2>0$ and $0$ (the latter, with multiplicity
$K-2$). Thus, for the Gaussian channel, $E_0(\bq)$ is also concave on the
simplex. On the other hand, one can easily find channels for which $D$ is not in
$\calD(K)$ and then $E_0(\bq)$ is not concave.

\section{Main Result}

Consider the following model of the FSC with an input--dependent state, which
is defined as follows:
\begin{equation}
\label{model}
P(\by|\bx)=\prod_{t=1}^n p(y_t|x_t,s_t),
\end{equation}
where the state $s_t\in\calS$ evolves recursively, in response to the channel
input, according to
\begin{equation}
\label{nextstate}
s_{t+1}=f(s_t,x_t),~~~t=1,2,\ldots,n-1,
\end{equation}
$f:\calS\times\calX\to\calS$ being a given {\it next--state function} and
$s_1$ is an arbitrary initial state.
It is assumed that the set of states $\calS$ has a finite cardinality, that
is, $S=|\calS|< \infty$, hence the qualifier ``finite--state''.

For the direct part of our coding theorem below, 
it is further assumed that the finite--state machine $f$ is irreducible,
namely, for every pair of states $s,s'\in\calS$,
there exists a finite string $x_1,x_2,\ldots,x_\ell\in\calX$ ($\ell \le S$) that
leads the machine from state $s$ to state $s'$.
Moreover, we assume that the finite--state machine formed by two
independent copies of $f$, that is,
the finite--state machine $(s_{t+1},s_{t+1}')=(f(s_t,x_t),f(s_t',x_t'))$,
is irreducible as well. For convenience, we henceforth refer to this assumption as {\it double
irreducibility}. 

For the converse part, we need a different assumption: we assume that there
exists a state $\sigma\in\calS$ and a positive integer $r$ such that for every
$s\in\calS$, there exists a path of length $r$, $x_1,x_2,\ldots,x_r$, that
takes the finite--state machine from state $s$ to state $\sigma$ (note that
by this definition, $r$ should be independent of $s$). We henceforth refer to
this assumption as {\it uniform approachability}. For example, if $\sigma$ has
a self--transition, this assumption is clearly satisfied. 

Before we present our main theorem, we first make a few simple observations.
Without loss of generality, we will take it for granted that the current state
$s_t$ contains the full information for recovery of $x_{t-1}$, that is, there
exists a deterministic function $g:\calS\to \calX$ such that 
\begin{equation}
\label{recover}
g(s_t)=x_{t-1}. 
\end{equation}
To justify the phrase ``without loss of generality'',
we note that for any given channel of the form (\ref{model}) and any given next--state
function $f$,
one can always artificially add the conditioning on $x_{t-1}$ in each
factor on the right--hand side (r.h.s.) of eq.\ (\ref{model}), that is,
represent the model as
\begin{equation}
\label{model1}
P(\by|\bx)=\prod_{t=1}^n p(y_t|x_t,x_{t-1},s_t),~~~t=1,2,\ldots,n-1,
\end{equation}
with some arbitrary definition of $x_0\in\calX$, and then re--define the state
as $\sigma_t=(s_t,x_{t-1})$. Having done this, we are back to the form
(\ref{model}), where:
(i) $s_t$ is replaced by $\sigma_t$, (ii) $\sigma_t$ evolves recursively in response
to $\{x_t\}$, using its own next--state function, 
and (iii) $x_{t-1}$ is recoverable from $\sigma_t$ simply because it includes
$x_{t-1}$ as a
component.\footnote{A simple important special case
where the assumption $x_{t-1}=g(s_t)$ 
is trivially satisfied, even without this modification, is the case where
$s_t=(x_{t-k},x_{t-k+1},\ldots,x_{t-1})$ ($k$ -- positive integer), which is
simply a shift register fed by $\{x_t\}$. This is the relevant case for the ISI
channel with a finite impulse response. In this case, the corresponding
finite--state machine also satisfies the double
irreducibility assumption and the uniform approachability assumption (for
example, the zero--state as a self--transition).}

Once the assumption (\ref{recover}) has been accepted, we
have the following simple equalities:
\begin{equation}
p(y_t|x_t,s_t)=p(y_t|x_t,s_t,s_{t+1})=p(y_t|s_t,s_{t+1}),
\end{equation}
where the first equality is due to the fact that $s_{t+1}$ is uniquely
determined by $x_t$ and $s_t$ (using $f$), and the second equality is because
in the presence of $s_{t+1}$, the conditioning on $x_t$ is redundant since
$x_t$ is determined by $s_{t+1}$ (using $g$). The mapping between $(x_t,s_t)$
and $(s_t,s_{t+1})$ is obviously one--to--one.
Thus, instead of modeling the
channel by the parameters $\{p(y|x,s),~x\in\calX,~s\in\calS\}$, one might as
well model it by the
parameters $\{p(y|s,s_+),~s,s_+\in\calS\}$, and think of the state sequence as the channel
input. Note that, in this parametrization, not all $S^2$ state pairs $(s,s_+)$ are necessarily feasible,
but only those
that are related by the equation 
\begin{equation}
s_+=f(s,g(s_+)), 
\end{equation}
in view of eqs.\
(\ref{nextstate}) and (\ref{recover}).
The number $L$ of feasible pairs
$\{(s,s_+):~s_+=f(s,g(s_+))\}$ cannot exceed $K\cdot S$, where $K$ denotes the size
of the input alphabet $\calX$, as before. An FSC with input--dependent states
is, therefore, completely defined by the functions $f$ and $g$, and the
parameters $\{p(y|s,s_+)\}$. Accordingly, we shall henceforth denote an
FSC by the
notation $[\{p(y|s,s_+)\},f,g]$.
Let us denote the Bhattacharyya distance between two state
pairs, $(s,s_+)$ and $(s',s_+')$ by
\begin{equation}
\label{bhat}
d_{\mbox{\tiny
B}}(s,s_+;s',s_+')=-\ln\left[\sum_{y\in\calY}\sqrt{p(y|s,s_+)p(y|s',s_+')}\right].
\end{equation}
The matrix $D$ of all Bhattacharyya
distances (\ref{bhat}) is, of course, of dimension $L\times L$.

We now redefine $\calQ_\Gamma$ to be the class of joint distributions
$\{q(s,s_+)\}$ of state
pairs that satisfy the following conditions:
\begin{enumerate}
\item For every state pair $(s,s_+)$: $q(s,s_+)> 0$ implies $s_+=f(s,g(s_+))$.
\item $\bq$ has equal marginals, i.e.,
$\sum_{\ts\in\calS}q(s,\ts)=\sum_{\ts\in\calS}q(\ts,s)\dfn\pi(s)$ for
every $s\in\calS$. 
\item All states in $\calS_+\dfn\{s:~\pi(s)> 0\}$ are fully connected, i.e.,
for every $s,s'\in\calS_+$, there exists a path $s=s_1\to s_2\to \ldots \to
s_m=s'$ (with $m\le|\calS_+|$), such that $q(s_i,s_{i+1})> 0$ for all $i=1,2,\ldots,m-1$.
\item The marginal $\bpi=\{\pi(s),~s\in\calS_+\}$ satisfies the input
constraint $\sum_{s\in\calS_+}\pi(s)\phi[g(s)]\le\Gamma$.
\end{enumerate}
Consider again the definitions of $\calE_0^+$ and $\calE_0^-$ 
as in (\ref{e0def}), but this time, with an FSC, rather
than a DMC, in mind. Also, $\calE_0^+(\Gamma)$ and $\calE_0^-(\Gamma)$ will be defined in the same way,
except that here, $P_e(n,R)$ is redefined as the minimum error probability across
all codes that satisfy the input constraint (\ref{inputconstraint})
for each codeword.
Accordingly, our new definition of $E_0(\bq)$ is
\begin{equation}
E_0(\bq)\dfn\sum_{s,s_+,s',s_+'}q(s,s_+)q(s',s_+')d_{\mbox{\tiny
B}}(s,s_+;s',s_+'),
\end{equation}
and once again, $\overline{E}_0(\bq)$ is the UCE of $E_0(\bq)$.
Considering the analogous extension of the r.h.s.\ of eq.\ (\ref{uce}),
here the time--sharing variable $u$ should take on values in an alphabet whose
size need not exceed $L$.
We are now ready to state our main theorem.

\begin{theorem}
Consider the FSC $[\{p(y|s,s^+)\},f,g]$, with
the input constraint (\ref{inputconstraint}). 
If the uniform approachability assumption is met,
\begin{equation}
\calE_0^+(\Gamma)\le\max_{\bq\in\calQ_\Gamma}\overline{E}_0(\bq).
\end{equation}
If $f$ is doubly irreducible, 
\begin{equation}
\calE_0^-(\Gamma)\ge\max_{\bq\in\calQ_\Gamma}\overline{E}_0(\bq).
\end{equation}
Consequently, if both assumptions hold,
\begin{equation}
\label{singletter}
\calE_0^+(\Gamma)=\calE_0^-(\Gamma)=\max_{\bq\in\calQ_\Gamma}\overline{E}_0(\bq).
\end{equation}
\end{theorem}

The remaining part of this section is devoted to the proof of Theorem 1.

{\it Proof.}
The proof is divided into two parts -- 
the direct part, asserting that
\begin{equation}
\calE_0^-(\Gamma)\ge \max_{\bq\in\calQ_\Gamma}\overline{E}_0(\bq),
\end{equation}
and the converse part, which tells that
\begin{equation}
\calE_0^+(\Gamma)\le
\max_{\bq\in\calQ_\Gamma}\overline{E}_0(\bq).
\end{equation}

Beginning with the direct part, to fix ideas, consider first the case where
$E_0(\bq)$ is concave and then $\overline{E}_0(\bq)=E_0(\bq)$.
Let $\bq^*$ be an\footnote{We
refer to {\it an} achiever, rather than {\it the} achiever, because 
for a general matrix $D$, the
maximum may be achieved by more than one distribution $\bq$.}
achiever of the $\max_{\bq\in\calQ_\Gamma} E_0(\bq)$.
For convenience, let us
assume\footnote{If this is not the case, one can slightly alter
$\bq^*$ with an arbitrarily small degradation in $E_0(\bq)$.}
that $q^*(s,s_+)\ge q_{\min}> 0$ for all state pairs for
which $s_+=f(s,g(s_+))$, thus $\calS_+=\calS$.
Consider an oriented multi--graph $G$ having
a total of $n$ arcs (edges) and $|\calS_+|$ vertices, labeled by the members of $\calS_+$. 
For every ordered pair $(s,s_+)$, let $G$ contain\footnote{We are
assuming, without essential loss of generality, 
that $\{nq^*(s,s_+)\}$ are all integers. If this is not the case,
$q^*(s,s_+)$ can be
approximated arbitrarily closely, for large $n$, by rational numbers with
denominator $n$.} $nq^*(s,s_+)$ arcs stemming from
vertex $s$ and ending at vertex $s_+$. 

From the construction in \cite[p.\ 433]{DLS81}, we learn that given such a directed multigraph
$G$, there exist (exponential many) state sequences of length $n$,
$\bs=(s_1,s_2,\ldots,s_n)$, with $s_1=f(s_n,g(s_1))$, that are identified with
various Eulerian circuits\footnote{An
Eulerian circuit is a walk on a graph, starting an ending at the same vertex,
where each arc is used exactly once.} on $G$. In other words, 
there exist many sequences $\bs$ with the property that the number of
transitions from $s_t=s$ to $s_{t\oplus 1}=s_+$ is exactly $nq^*(s,s_+)$, where
$\oplus$ denotes addition modulo $n$, that is, we adopt the cyclic
convention that $s_n$ is followed by $s_1$ (hence the requirement
$s_1=f(s_n,g(s_1))$). The validity of this statement is based on properties of
$G$ that are guaranteed by the definition of the class $\calQ_\Gamma$ to which
$\bq^*$ belongs (see, in particular,
properties 3 and 4 in \cite[p.\ 433]{DLS81}, which are reflected in items 2
and 3 in the definition of $\calQ_\Gamma$). For convenience, we make the
convention that the initial state $s_1$ is always a certain fixed member
$\sigma$ of $\calS$.

Let $\calT_n(\bq^*)$ be the set
of all state sequences $\{\bs\}$ with the properties described in the previous
paragraph, that is, the so called {\it Markov
type} associated with $\bq^*$ (see, e.g., \cite{DLS81},
\cite[Subsection VII.A]{Csiszar98} and references therein).
Let $M$ be a fixed (independent of $n$) positive integer and
consider an independent random selection of $2M-1$ members from
$\calT_n(\bq^*)$,
each one under the uniform distribution across $\calT_n(\bq^*)$, i.e.,
\begin{equation}
\Pi(\bs)=\left\{\begin{array}{ll}
\frac{1}{|\calT_n(\bq^*)|} & \bs\in\calT_n(\bq^*)\\
0 & \mbox{elsewhere}\end{array}\right.
\end{equation}
Let $\bs_1,\bs_2,\ldots,\bs_{2M-1}$ be the resulting randomly chosen state
sequences. We can think of this collection as a random code
for the channel 
\begin{equation}
\label{s2ychannel}
P(\by|\bs)\dfn\prod_{t=1}^n p(y_t|s_t,s_{t\oplus 1}).
\end{equation}
We next apply an expurgation process 
(see, e.g., \cite[Subsection 5.7]{Gallager68}, \cite[Subsection 3.3]{VO79}), 
which guarantees 
that there exists a sub-code
of size $M$ for which each each codeword
contributes a conditional error probability that does not
exceed $(2\overline{P_{\mbox{\tiny e|m}}^{1/\rho}})^\rho$, where $\rho$ is 
an arbitrary positive real, and $\overline{P_{\mbox{\tiny e|m}}^{1/\rho}}$ is
the expectation of $P_{\mbox{\tiny e|m}}^{1/\rho}$ under the above
defined ensemble. Therefore, within this sub-code,
\begin{equation}
\label{upper}
\max_{1\le m\le M} P_{\mbox{\tiny e}|m}\le
\left\{4M\sum_{\bs,\bs'}\Pi(\bs)\Pi(\bs')
\left[\sum_{\by}\sqrt{P(\by|\bs)P(\by|\bs')}\right]^{1/\rho}\right\}^\rho,
\end{equation}
and consequently,
\begin{eqnarray}
\label{expurg}
& &\limsup_{n\to\infty}\frac{\ln\left[\max_{1\le m\le M} P_{\mbox{\tiny
e}|m}\right]}{n}\nonumber\\
&\le&\liminf_{\rho\to\infty}\limsup_{n\to\infty}\frac{1}{n}\ln\left(
\left\{4M\sum_{\bs,\bs'}\Pi(\bs)\Pi(\bs')
\exp\left[-\frac{1}{\rho}\sum_{t=1}^nd_{\mbox{\tiny B}}(s_t,s_{t\oplus 1};s_t',s_{t\oplus
1}')\right]\right\}^{\rho}\right)\nonumber\\
&\lea&-\sum_{s,s_+,s',s_+'}q^*(s,s_+)q^*(s',s_+')d_{\mbox{\tiny
B}}(s,s_+;s',s_+')\nonumber\\
&=&-E_0(\bq^*)=-\max_{\bq\in\calQ_{\Gamma}}E_0(\bq),
\end{eqnarray}
where the inequality marked by (a) is justified by using
the method of types for Markov chains (\cite[Subsection VII.A]{Csiszar98} and references therein),
and on the basis of the double irreducibility assumption (see Appendix for the details).

Finally, let $\{\bs_1,\ldots,\bs_M\}$
be a sub--code
with the property $\max_{1\le m \le M}P_{\mbox{\tiny e}|m}
\le e^{-n[E_0(\bq^*)-o(n)]}$
(where the indices $1,2,\ldots,M$ are after possible relabeling).
Then each $n$-tuple $\bs_m=(s_{m,1},\ldots,s_{m,n})$, 
$m=1,2.\ldots,M$, uniquely determines a corresponding
codeword $\bx_m=(x_{m,1},\ldots,x_{m,n})$ according to $x_{m,t}=g(s_{m,t\oplus
1})$, $t=1,2,\ldots,n$, which obviously satisfies
the input constraint (\ref{inputconstraint}), and so, the actual code for the given channel is
$\calC=\{\bx_1,\ldots,\bx_M\}$. 
This completes the proof of the direct part the for case where $E_0(\bq)$ is
concave. 

To complete the proof of the direct part for the general case, we repeat the very same
construction, but now, we combine it with time sharing. In particular, consider the more explicit
form of $\overline{E}_0(\bq^*)$ as
\begin{equation}
\label{uce1}
\overline{E}_0(\bq^*)=\max_{\bw,\bV}\sum_{u\in\calU}w(u)E_0[v(\cdot,\cdot|u)],
\end{equation}
where $\bw=\{w(u),~u\in\calU\}$ is a probability assignment on $u$,
$\bV=\{v(s,s_+|u),~s,s_+\in\calS,~s_+=f(s,g(s_+)),~u\in\calU\}$ is
a set of probability assignments on state pairs given $u$, and the
maximum is over all pairs $\{(\bw,\bV)\}$ such that
$\sum_{u\in\calU}w(u)V(s,s_+|u)=q^*(s,s_+)$.
Let $\bw^*$ and
$\bV^*$ be achievers of the maximum on the r.h.s.\ of (\ref{uce1}).
For each codeword, the block of length $n$ is divided into $|\calU|$ segments, each one of length
$nw^*(u)$, labeled by $u\in\calU$. Specifically, for every $m=1,2,\ldots,M$, we proceed as
follows. For every $u\in\calU$, select, independently at random, a member from 
$\calT_{nw^*(u)}[v^*(\cdot,\cdot|u)]$, as the $u$--th segment of the state sequence
associated with codeword, which is
concatenated to all previous segments. Now, after expurgation
of such randomly selected code, a
straightforward extension of the derivation in (\ref{upper}) and
(\ref{expurg}) would yield
an error exponent of
$\sum_{u\in\calU}w^*(u)E_0[v^*(\cdot,\cdot|u)]=\overline{E}_0(\bq^*)$.
Note that there is no need to worry about tailoring consecutive segments of
the state sequence, because by our convention, all segments begin and end at state $\sigma$.
This completes the proof of the direct part.

Moving on to the converse part,
let $\calC$ be an arbitrary rate--$\epsilon$ code ($\epsilon > 0$,
infinitesimally small) of length $n$,
that satisfies the input constraint (\ref{inputconstraint}) for each
codeword. Consider the transformation of each codeword $\bx_m$ in $\calC$ into a
state sequence $\bs_m$, according to the recursion $s_{m,t+1}=f(s_{m,t},x_{m,t})$,
$t=1,2,\ldots,n-1$, $m=1,2,\ldots,M=e^{n\epsilon}$, where $s_{m,1}=\sigma$,
which is a uniformly approachable state, and all codewords are extended (if
needed) to be of length $n'= n+r$ (complying with the same recursion also for
$t=n,n+1,\ldots,n'-1$), such that $f(s_{n'},x_{n'})=
\sigma$, which is possible by the uniform approachability assumption. 
This extension of the codewords can only decrease the probability of
error, so any lower bound on the error probability of the modified code
is also a lower bound for the original code. The price of this extension is a possible
increase in the average cost, but by no more than $r\cdot\max_x\phi(x)/n\dfn
c/n$, which is vanishing as
$n$ grows without bound, since $r$ depends only on $f$, but not on $n$.

For the sake of convenience, we
denote the new block length by $n$ again, rather than $n'$.
Consider now the resulting collection of state sequences, $\{\bs_1,\bs_2,\ldots,\bs_M\}$,
which can be considered as a code for the channel (\ref{s2ychannel}).
Obviously, each $\bs_m$ belongs to some Markov type $\calT_n(\bq)$ 
where $\bq\in\calQ_{\Gamma+c/n}$.
Since the number of distinct Markov types cannot exceed
$(n+1)^{S^2}$, then at least $(n+1)^{-S^2}e^{n\epsilon}$ `codewords' 
must belong to the same Markov type $\calT_n(\bq)$.
Obviously, the probability of error of the
original given code (after the extension) cannot be smaller than $(n+1)^{-S^2}$ times the
probability of error of the smaller code $\calC'=|\calC\cap\calT_n(\bq)|$.
Thus, any upper bound on the error exponent of $\calC'$ is also an upper bound
on the error exponent of the original code, and so, from this point onward we
may assume that all codewords are of the same Markov type $\calT_n(\bq)$,
$\bq\in\calQ_{\Gamma+c/n}$. 

Now, the channel (\ref{s2ychannel}) is obviously memoryless w.r.t.\ pairs of
consecutive states $\{(s_t,s_{t\oplus 1})\}$, and 
we can therefore invoke the proof
of Theorem 4 in \cite{SGB67b} for memoryless channels.
Combining eqs.\ (1.12), (1.36), (1.40), (1.42), (1.43) and (1.53) of
\cite{SGB67b} (with $K$ of \cite{SGB67b} being replaced by $L$, in our notation), 
we learn that
\begin{equation}
-\frac{\ln P_e(\epsilon,n)}{n}\le \frac{1}{M^2}\sum_{t=1}^n\sum_{s,s_+,s',s_+'}
M_t(s,s_+)M_t(s',s_+')d_{\mbox{\tiny
B}}(s,s_+;s',s_+')+o(n),
\end{equation}
where $M_t(s,s_+)$ is the number of codewords in (a subset of) $\calC'$ such that
$(s_{m,t},s_{m.t\oplus 1})=(s,s_+)$ and $o(n)$ is a term that tends to zero
as $n\to\infty$. It now readily follows that
\begin{eqnarray}
\calE_0^+(\Gamma)&=&\lim_{\epsilon\downarrow 0}\limsup_{n\to\infty}\left[-\frac{\ln
P_e(\epsilon,n)}{n}\right]\\
&\le&\limsup_{n\to\infty}\left[\frac{1}{n}\sum_{t=1}^nE_0\left(\frac{M_t(\cdot,\cdot)}{M}\right)\right]\\
&\le&\limsup_{n\to\infty}\left[\frac{1}{n}\sum_{t=1}^n
\overline{E}_0\left(\frac{M_t(\cdot,\cdot)}{M}\right)\right]\\
&\le&\limsup_{n\to\infty}\overline{E}_0\left(\frac{1}{n}\sum_{t=1}^n
\frac{M_t(\cdot,\cdot)}{M}\right)\\
&=&\limsup_{n\to\infty}\overline{E}_0(\bq)\\
&\le&\limsup_{n\to\infty}\sup_{\bq\in\calQ_{\Gamma+c/n}}\overline{E}_0(\bq)\\
&=&\sup_{\bq\in\calQ_\Gamma}\overline{E}_0(\bq).
\end{eqnarray}
This completes the proof of the converse part, and hence also the proof of
Theorem 1. 

\section{The Gaussian Channel with ISI}

In this section, we consider the important special case of the Gaussian
channel with ISI. Our objective is to provide more explicit results, which are
available thanks to the facts that: (i) the finite--state machine $f$ is simple,
and more importantly, and (ii) the Bhattacharyya distance is proportional to the
Euclidean distance, for which $E_0(\bq)$ is concave, and hence the operator
$\mbox{UCE}\{\cdot\}$ becomes redundant.

The Gaussian ISI channel 
is defined by
\begin{equation}
\label{isichannel}
y_t=\sum_{i=0}^kh_ix_{t-i}+w_t,
\end{equation}
where $\{w_t\}$ is Gaussian white noise with zero mean, variance
$\sigma^2$, and is independent of the channel input, $\{x_t\}$.
Here, $\{h_i\}_{i=0}^k$ are the ISI channel coefficients. Obviously, the
state of the channel, in this case, is given by the contents a shift register
of length $k$, fed by the input, i.e.,
$s_t=(x_{t-k},x_{t-k+1},\ldots,x_{t-1})\dfn x_{t-k}^{t-1}$, and the corresponding next--state
function $f$ is doubly irreducible and uniformly approachable.
The channel input power is limited to $\Gamma$, that is, 
the input constraint (\ref{inputconstraint}) is imposed with the cost function
$\phi(x)=x^2$.

First, a straightforward calculation of the Bhattacharyya distance for the
Gaussian ISI channel (\ref{isichannel}) yields
\begin{eqnarray}
d_{\mbox{\tiny B}}(s_t,s_{t\oplus 1};\ts_t,\ts_{t\oplus 1})&=&
d_{\mbox{\tiny B}}(x_{t-k}^t,\tx_{t-k}^t)\nonumber\\
&=&\frac{1}{8\sigma^2}\left(\sum_{i=0}^kh_tx_{t-i}-\sum_{i=0}^kh_t\tx_{t-i}\right)^2.
\end{eqnarray}
Therefore,
\begin{eqnarray}
E_0(\bq)&=&\frac{1}{8\sigma^2}
\sum_{x_0^k,\tx_0^k}q(x_0^k)q(\tx_0^k)
\left(\sum_{i=0}^kh_ix_{k-i}-\sum_{i=0}^kh_i\tx_{k-i}\right)^2\\
&=&\frac{1}{4\sigma^2}\left[\sum_{x_0^k}q(x_0^k)
\left(\sum_{i=0}^kh_ix_{k-i}\right)^2-\left(\sum_{x_0^k}q(x_0^k)
\sum_{i=0}^kh_ix_{k-i}\right)^2\right]\nonumber\\
&=&\frac{1}{4\sigma^2}\sum_{x_0^k}q(x_0^k)
\left(\sum_{i=0}^kh_ix_{k-i}\right)^2-
\frac{1}{4\sigma^2}\left(\sum_{i=0}^kh_i\sum_{x_{k-i}}q(x_{k-i})x_{k-i}\right)^2\nonumber\\
&=&\frac{1}{4\sigma^2}\sum_{i=0}^k\sum_{j=0}^kh_ih_j\sum_{x_0x_{|i-j|}}q(x_0,x_{|i-j|})x_0x_{|i-j|}
-\frac{1}{4\sigma^2}\left(\sum_{i=0}^kh_i\sum_{x_0}q(x_0)x_0\right)^2.
\end{eqnarray}
The above expression should be maximized subject to a set of constraints that
reflect the fact that $q(x_0^k)$ stems from an empirical distribution (of each
codeword), i.e.,
the marginals of $(x_{i_1},x_{i_2},\ldots,x_{i_l})$ ($l\le k$)
depend on the indices $i_1,i_2,\ldots,i_l$ only via the differences $i_2-i_1,
i_3-i_2,\ldots, i_l-i_{l-1}$. An additional constraint is, of course, the
power constraint $\sum_{x_0}q(x_0)x_0^2\le \Gamma$. Since the objective
function is concave in $\bq$ and the constraints are linear, this is, in
principle, a standard
convex programming problem.

It would be insightful to examine now the behavior in the case where $\{x_t\}$ 
takes on continuous values on the real line.
In this case, in the limit of large $n$,
the last expression reads, in the frequency domain, as follows:
\begin{equation}
E_0(\bq)=\frac{1}{4\sigma^2}\left[\frac{1}{2\pi}\int_{-\pi}^{+\pi}
S_x(e^{i\omega})|H(e^{i\omega})|^2\mbox{d}\omega-\bar{X}^2|H(e^{i
0})|^2\right],
\end{equation}
where $H(e^{i\omega})$ ($i=\sqrt{-1}$) is the frequency response (the Fourier
transform) associated with impluse response $\{h_i\}_{i=0}^k$, 
$S_x(e^{i\omega})$ is power spectrum of an underlying stationary process
$\{X_t\}$, and
$\bar{X}$ is the DC component of $\{X_t\}$. In
other words,
we think of the input power spectrum as
\begin{equation}
S_x(e^{i\omega})=S_x'(e^{i\omega})+2\pi\bar{X}\delta(\omega),~~~~-\pi\le\omega
<\pi
\end{equation}
where $S_x'(e^{i\omega})$ does not include a Dirac delta function at the
origin. We can now express the zero--rate exponent as
\begin{equation}
E_0(\bq)=\frac{1}{4\sigma^2}\cdot\frac{1}{2\pi}\int_{-\pi}^{+\pi}
S_x'(e^{i\omega})|H(e^{i\omega})|^2\mbox{d}\omega,
\end{equation}
which should be maximized under the power constraint 
$$\frac{1}{2\pi}\int_{-\pi}^{+\pi}S_x'(e^{i\omega})\mbox{d}\omega+\bar{X}^2\le \Gamma.$$
It is now obvious that any non--zero value of $\bar{X}$ is just a waste, at
the expense of the available
power, which does not contribute to $E_0(\bq)$, and the best input spectrum
is that of a sinusoidal process at the frequency $\omega_0$ that
maximizes the amplitude response $|H(e^{i\omega})|$. If $\omega_0=0$, this
means a DC process, which strictly speaking, contradicts our conclusion that the DC component
should vanish. In this case, one can approach the maximum achievable exponent
by a sinusoidal waveform of an arbitrarily low frequency, so that the response
is close as desired to the maximum. Thus, the maximum achievable exponent when
$\calX=\reals$ is given by
\begin{equation}
\sup_{\bq\in\calQ_\Gamma}E_0(\bq)=\frac{\Gamma}{4\sigma^2}\cdot\max_\omega|H(e^{i\omega})|^2.
\end{equation}
To create $M$ orthogonal codewords, one can generate each one with a
slightly different frequency in the vicinity of $\omega_0$. 
This is, of course, an upper bound also for any discrete--alphabet input.

It would be interesting now to have also a lower bound on the achievable
zero--rate exponent for a given finite--alphabet size $K$.
To this end, we will analyze the behavior for a specific class of input
signals. When the finite input alphabet corresponds to the $K$ quantization levels of a
uniform quantizer $Q(\cdot)$, i.e., $\{\pm (i-1/2)\Delta,~i=1,2,\ldots,K/2\}$ ($K$ even), and
$\Delta$ is reasonably small, it is natural, in view of the above, to consider the quantized sinusoid
as an input signal
$x_t=Q[A\sin(\omega_0t+\phi)]$, where $A\le (K-1)\Delta/2$ is chosen to meet
the input power constraint, $\sum_tQ^2[A\sin(\omega_0t+\phi)]\le n\Gamma$.
Obviously, the smaller is $\Delta$ (i.e., the larger is $K$ for a given $A$),
the smaller is the loss compared to the clean (unquantized) sinusoid.
We next examine this loss.

Let $e_t=Q[A\sin(\omega_0t+\phi)]-A\sin(\omega_0t+\phi)$ designate the
quantization error signal.
Then,
\begin{eqnarray}
\Gamma&=&\frac{1}{n}\sum_{t=1}^nQ^2[A\sin(\omega_0t+\phi)]\\
&=&\frac{1}{n}\sum_{t=1}^n[A\sin(\omega_0t+\phi)+ e_t]^2\\
&=&\frac{A^2}{2}+\frac{2}{n}\sum_{t=1}^nAe_t\sin(\omega_0t+\phi)+\frac{1}{n}\sum_{t=1}^ne_t^2\\
&\to&\frac{A^2}{2}+2R_{xe}(0)+R_{ee}(0),
\end{eqnarray}
where we define
\begin{equation}
R_{ee}(\ell)=\lim_{n\to\infty}\frac{1}{n}\sum_{t=1}^ne_te_{t+\ell}
\end{equation}
and
\begin{equation}
R_{xe}(\ell)=\lim_{n\to\infty}\frac{A}{n}\sum_{t=1}^ne_{t+\ell}\sin(\omega_0t+\phi).
\end{equation}
Denoting $H_{\max}^2=|H(e^{i\omega_0})|^2$, we now have:
\begin{eqnarray}
& &\lim_{n\to\infty}\frac{1}{n}\sum_{t=1}^n\left[\sum_{\ell=0}^kh_\ell
x_{t-\ell}\right]^2\\
&=&\lim_{n\to\infty}\frac{1}{n}\sum_{t=1}^n\left[\sum_{\ell=0}^kh_\ell(A\sin[\omega_0(t-\ell)+
\phi]+e_{t-\ell})\right]^2\\
&=&\frac{A^2}{2}H_{\max}^2+\sum_{\ell=0}^k\sum_{j=0}^kh_\ell h_j[R_{xe}(\ell-j)+R_{xe}(j-\ell)+R_{ee}(\ell-j)]\\
&=&[\Gamma-2R_{xe}(0)-R_{ee}(0)]H_{\max}^2+\sum_{\ell=0}^k\sum_{j=0}^kh_\ell
h_j[R_{xe}(\ell-j)+R_{xe}(j-\ell)+R_{ee}(\ell-j)]\\
&=&\Gamma H_{\max}^2-\Lambda
\end{eqnarray}
where $\Lambda$ is the loss due to quantization, i.e.,
\begin{equation}
\Lambda=[2R_{xe}(0)+R_{ee}(0)]H_{\max}^2-
\sum_{\ell=0}^k\sum_{j=0}^kh_\ell
h_j[R_{xe}(\ell-j)+R_{xe}(j-\ell)+R_{ee}(\ell-j)].
\end{equation}
For the case where $\omega_0$ is irrational,
one can find in \cite[eqs.\ (44), (45), (51)]{Gray90} all the relevant joint second order
statistics needed here. In particular, for the sinuoidal input under
discussion,
\begin{equation}
R_{ee}(\ell)=\sum_{m=-\infty}^\infty \varepsilon_m \exp\{2\pi i\ell
\lambda_m\}=\sum_{m=-\infty}^\infty \varepsilon_m\cos(2\pi \ell\lambda_m),
\end{equation}
where $\lambda_m=\left<(2m-1)\omega_0/2\pi\right>$, and
\begin{equation}
\varepsilon_m=\left[\frac{\Delta}{\pi}\sum_{\ell=1}^\infty\frac{J_{2m-1}(2\pi
\ell A/\Delta)}{\ell}\right]^2,
\end{equation}
$J_m(z)$ being the $m$--th coefficient in the Fourier series expansion of
the periodic function $\exp(iz\sin s)$, as a function of $s$, and
\begin{equation}
R_{xe}(\ell)=A\Delta\cos(\omega_0\ell)\sum_{m=1}^\infty \frac{J_1(2\pi
mA/\Delta)}{m}\dfn AB\cos(\omega_0\ell).
\end{equation}
We therefore obtain
\begin{eqnarray}
\Lambda&=&\left[2AB+\sum_{m=-\infty}^\infty \varepsilon_m\right]H_{\max}^2-2AB
\sum_{\ell=0}^k\sum_{j=0}^kh_\ell
h_j\cos[\omega_0(\ell-j)]-\nonumber\\
& &\sum_{m=-\infty}^\infty
\varepsilon_m\sum_{\ell=0}^k\sum_{j=0}^kh_\ell h_j\cos[2\pi(\ell-j)\lambda_m]\\
&=&\left[2AB+\sum_{m=-\infty}^\infty \varepsilon_m\right]H_{\max}^2-2H_{\max}^2AB
-\sum_{m=-\infty}^\infty
\varepsilon_m|H(e^{2\pi i\lambda_m})|^2\\
&=&\sum_{m=-\infty}^\infty \varepsilon_m[H_{\max}^2-|H(e^{2\pi i\lambda_m})|^2].
\end{eqnarray}
This expression is intuitively appealing: each term is the loss due to
spectral term of $\{e_t\}$ that is 
in a non--optimal frequency (higher order harmonic) $\lambda_m$,
where the power gain is $|H(e^{2\pi i\lambda_m})|^2$,
rather than the optimal frequency $\omega_0$, where the power gain is
$|H(e^{2\pi i\omega_0})|^2=H_{\max}^2$.
Thus, to summarize, the exponent of the finite--alphabet case is upper bounded
by $\Gamma H_{\max}^2/(4\sigma^2)$ and lower bounded by
$(\Gamma H_{\max}^2-\Lambda)/(4\sigma^2)$, where it should be kept in mind that
$\Lambda$ depends on the ratio $A/\Delta\le (K-1)/2$ via $\{\varepsilon_m\}$.
In \cite[eq.\ (50)]{Gray90}, there is a more explicit expression for
$\varepsilon_m$. As $K$ increases, the loss $\Lambda$ decreases, essentially inverse
proportionally to $K^2$,

On a related note, in the continuous--time version of the problem, where the
channel is an additive white Gaussian noise channel, 
without bandwidth constraints, but only a peak--power
constraint, a binary input $x_t\in\{-\sqrt{\Gamma},+\sqrt{\Gamma}\}$ is as good as any
$x_t\in[-\sqrt{\Gamma},+\sqrt{\Gamma}]$ since the filter response to the latter can be
approximated arbitrarily closely using binary inputs, as is shown in
\cite{OWZ88}. In other words, when $\{x_t\}$ is not discretized in time, it
can be discretized in amplitude even as coursely as in binary quantization
without essential loss of optimality.

%\section*{Acknowledgements}

\section*{Appendix}
\renewcommand{\theequation}{A.\arabic{equation}}
    \setcounter{equation}{0}

{\it Justification of Inequality (a) in Equation (\ref{expurg}).}
We are interested in an exponential upper bound on the expression
\begin{equation}
\left\{\Pi(\bs)\Pi(\bs')
\exp\left[-\frac{1}{\rho}\sum_{t=1}^nd_{\mbox{\tiny B}}(s_t,s_{t\oplus
1};s_t',s_{t\oplus
1}')\right]\right\}^{\rho}.
\end{equation}
Using the method of types for Markov chains,
we find that the exponential rate of this quantity is of the exponential order
of $\exp\{-nZ(\rho)\}$, where
\begin{eqnarray}
Z(\rho)&=&\min_{w_{SS_+S'S_+'}}\left\{\rho[H_w(S_+|S)+
H_w(S_+'|S')-H_w(S_+,S_+'|S,S')]-\right.\nonumber\\
& &\left.\sum_{s,s_+,s',s_+'}w_{SS_+S'S_+'}(s,s_+,s',s_+')
d_{\mbox{\tiny B}}(s,s_+;s',s_+')\right\},
\end{eqnarray}
where $w_{SS_+S'S_+'}$ is a generic joint distribution of a dummy quadruple of
random variables $(S,S_+,S',S_+')$ over $\calS^4$, $H_w(\cdot|
\cdot)$ are various conditional entropies induced by $w_{SS_+S'S_+'}$, and the weighted
divergences are defined in the usual way. We note that since $w_{SS_+S'S_+'}$
is the empirical distribution of two pairs of consecutive states, it must
always satisfy the stationarity conditions
\begin{equation}
\label{stationary}
\sum_{s_1,s_2}w_{SS'}(s_1,s_2)w_{S_+S_+'|SS'}(s_3,s_4|s_1,s_2)=w_{SS'}(s_3,s_4)~~~~\forall
s_3,s_4.
\end{equation}
Since $\Pi(\bs)$ supports only members in $\calT_n(\bq^*)$, we also note that
\begin{eqnarray}
\sum_{s',s_+'}w_{SS_+S'S_+'}(s,s_+,s',s_+')&=&q^*(s,s_+)\\
\sum_{s,s_+}w_{SS_+S'S_+'}(s,s_+,s',s_+')&=&q^*(s',s_+').
\end{eqnarray}
Now, let is denote
\begin{equation}
\Delta(w_{SS_+S'S_+'})=H_w(S_+|S)+H_w(S_+'|S')-H_w(S_+,S_+'|S,S')
\end{equation}
and note that $\Delta(w_{SS_+S'S_+'})\ge 0$ with equality iff $w_{SS_+S'S_+'}$
satisfies $w_{SS_+S'S_+'}(s,s_+,s',s_+')=w_{SS'}(s,s')q^*(s_+|s)q^*(s_+'|s')$, where
$q^*(s_+|s)\dfn q^*(s,s_+)/\pi^*(s)$.
Now, let $w_{SS_+S'S_+'}^\rho$ denote the minimizing $w_{SS_+S'S_+'}$ for a
given $\rho$. Considering a sequence $\rho_\ell\to\infty$, 
we have
\begin{eqnarray}
\label{limit}
\limsup_{\ell\to\infty}Z(\rho_\ell)&=&
\limsup_{\ell\to\infty}\min_{w_{SS_+S'S_+'}}\left[\rho_\ell
\cdot\Delta(w_{SS_+S'S_+'})-\right.\nonumber\\
& &\left.-\sum_{s,s_+,s',s_+'}w_{SS_+S'S_+'}(s,s_+,s',s_+')
d_{\mbox{\tiny B}}(s,s_+;s',s_+')\right]\nonumber\\
&=&\limsup_{\ell\to\infty}\left[\rho_\ell\cdot\Delta(w_{SS_+S'S_+'}^{\rho_\ell})-\right.\nonumber\\
& &\left.-\sum_{s,s_+,s',s_+'}w_{SS_+S'S_+'}^{\rho_\ell}(s,s_+,s',s_+')
d_{\mbox{\tiny B}}(s,s_+;s',s_+')\right]\nonumber\\
&\ge&-\liminf_{\ell\to\infty}\left[
\sum_{s,s_+,s',s_+'}w_{SS_+S'S_+'}^{\rho_\ell}(s,s_+,s',s_+')
d_{\mbox{\tiny B}}(s,s_+;s',s_+')\right].
\end{eqnarray}
As $\ell\to\infty$, there is a subsequence with indices $\{\ell_i\}$
that tends to the limit inferior in the last line of
(\ref{limit}), and within this subsequence, there is a sub--subsequence for
which $w_{SS_+S'S_+'}^{\rho_{\ell_i}}(s,s_+,s',s_+')$ converges\footnote{
This is true since the space of joint
distributions on a finite support is compact.}
to some limiting distribution of the form
of the form $w_\infty(s,s')q^*(s_+|s)q^*(s_+'|s')$, as otherwise,
the $\rho\cdot\Delta$ term would tend to infinity, and hence cannot achieve
the minimum, which is finite.
Thus,
\begin{equation}
\limsup_{\ell\to\infty}Z(\rho_\ell)\ge-\sum_{s,s_+,s',s_+'}
w_\infty(s,s')q^*(s_+|s)q^*(s_+'|s')d_{\mbox{\tiny B}}(s,s_+;s',s_+').
\end{equation}
Now, since $w_\infty(s,s')q^*(s_+|s)q^*(s_+'|s')$ is a limit of
empirical distributions of pairs of consecutive states, then, as mentioned in
(\ref{stationary}), it must satisfy 
\begin{equation}
\sum_{s,s'}w_\infty(s,s')q^*(s_+|s)q^*(s_+'|s')=w_\infty(s_+,s_+')~~~~\forall~s_+,s_+'.
\end{equation}
One solution to these equations is obviously
$w_\infty(s,s')=\pi^*(s)\pi^*(s')$, but since we have assumed double
irreducibility, then
the corresponding pair of independent Markov chains has a
unique stationary state distribution, which then must be $\pi^*(s)\pi^*(s')$.
Thus,
\begin{eqnarray}
\limsup_{\ell\to\infty}Z(\rho_\ell)&\ge&-\sum_{s,s_+,s',s_+'}
\pi^*(s)\pi^*(s')q^*(s_+|s)q^*(s_+'|s')d_{\mbox{\tiny
B}}(s,s_+;s',s_+')\nonumber\\
&=&-\sum_{s,s_+,s',s_+'}
q^*(s,s_+)q^*(s',s_+')d_{\mbox{\tiny B}}(s,s_+;s',s_+').
\end{eqnarray}

\end{document}